\documentstyle[epsfig,aps,prl,twocolumn]{revtex}
\begin{document}

\wideabs{
\title{Electron-phonon coupling induced pseudogap and the superconducting transition in Ba$_{0.67}$K$_{0.33}$BiO$_{3}$}

\author{A. Chainani,$^{1,2}$ T. Yokoya,$^{1}$ T. Kiss,$^{1}$ S. Shin,$^{1,3}$ T. Nishio,$^{4}$ and H. Uwe$^{4}$}

\address{
$^{1}$Institute for Solid State Physics, University of Tokyo, Kashiwa,Chiba 277-8581, Japan
}
\address{
$^{2}$Institute for Plasma Research, Gandhinagar 382 428, India
}
\address{
$^{3}$The Institute of Physical and Chemical Research(RIKEN), Sayo-gun, Hyogo 679-5143, Japan
}
\address{
$^{4}$Institute of Materials Science, University of Tsukuba, Tsukuba, Ibaraki 305-8573, Japan
}

\date{\today}

\maketitle

\begin{abstract}\\
\hspace{0.5cm}We study the single particle density of states (DOS) across the superconducting transition ({\it T$_{c}$} = 31 K) in single-crystal Ba$_{0.67}$K$_{0.33}$BiO$_{3}$ using ultrahigh resolution angle-integrated photoemission spectroscopy. The superconducting gap opens with a pile-up in the DOS, $\Delta$(5.3 K) = 5.2 meV and 2$\Delta$(0)/{\it k$_{B}$T$_{c}$} = 3.9. In addition, we observe a pseudogap below and above {\it T$_{c}$}, occurring as a suppression in intensity over an energy scale up to the breathing mode phonon($\sim$ 70 meV). The results indicate electron-phonon coupling induces a pseudogap in Ba$_{0.67}$K$_{0.33}$BiO$_{3}$ . 
\end{abstract}
}

From a variety of experiments and theory, it is clear that the high temperature cuprate superconductors belong to a special class of materials. Some of the important aspects of the cuprates are quasi two-dimensionality, short-range antiferromagnetic correlations, an anisotropic pseudogap in the normal phase, and d$_{x^{2}-y^{2}}$ symmetry of the superconducting gap[1,2]. In contrast, the perovskite series Ba$_{1-x}$K$_{x}$BiO$_{3}$ (BKBO) is three-dimensional, contains no transition metal ions, has no magnetic order but still exhibits the highest {\it T$_{c}$} known for an oxide other than the cuprates[3]. The superconductivity in the BKBO series was preceded by the analogous BaBi$_{1-x}$Pb$_{x}$O$_{3}$ (BPBO) series which shares the same parent BaBiO$_{3}$ (BBO)[4]. The role of electron-phonon coupling in the properties of the BKBO and BPBO series originates in the charge density wave (CDW) state of BBO.  

The parent BBO is expected to be a metal from band theory with a half-filled Bi 6s band, but the near perfect nesting possible in BBO causes  a three-dimensional CDW gap to open up in the DOS[4,5]. The CDW in BBO is coupled to the breathing mode phonon which is due to the contraction and expansion of oxygen octahedra surrounding neighbouring Bi ions. Subtitution with K in the Ba-site results in a semiconductor-metal transition at a critical x$_{c}$ $\sim$ 0.3. While the CDW state is weakened by substitution, resulting in a systematic lowering of the CDW energy upon doping, the breathing mode phonon is observed even in the metallic phase[6]. Optical conductivity studies[7] suggest remnant local CDW order in the superconducting compositions with x = 0.33 and 0.4. EXAFS measurements[8] also show that there are two types of Bi ions, with short- and long-bond nearest neighbour Bi-O distances. On increasing x, the short and long Bi-O distances become equivalent around x = 0.4. This picture is derived from the formally Bi$^{3+}$ and Bi$^{5+}$ ions constituting the charge disproportionation in BBO[4], though X-ray and resonant photoemission spectroscopy studies have not shown clear evidence for the same in BBO and the doped compounds[9,10]. The results are understood as due to very small charge transfer between the unequal Bi sites[5]. The crystal structure also evolves with doping and at room temperature, for x = 0.0 - 0.1 it is monoclinic, from x = 0.1 to 0.3 it is orthorhombic and above x = 0.3 up to 0.5( = the solubility limit), it is cubic[11].

In this work we study the electronic structure of Ba$_{1-x}$K$_{x}$BiO$_{3}$ with   x = 0.2, 0.33 and 0.46. For x = 0.2, the system is semiconducting ; for x = 0.33 we are across the critical concentration x$_{c}$ $\sim$ 0.3 and into the metallic phase with a T$_{c}$ of 31 K; x= 0.46 is overdoped with a {\it T$_{c}$} of $\sim$24 K. We have done a detailed study with the purpose of investigating the role of charge order (in the absence of magnetic order) on the electronic structure and superconducting transition in Ba$_{0.67}$K$_{0.33}$BiO$_{3}$. Our results show that the superconducting gap opens due to a pile up in the DOS with spectacular redistribution of spectral weight at low energy scales. Ba$_{0.67}$K$_{0.33}$BiO$_{3}$ belongs to the moderately strong electron-phonon coupling regime with 2$\Delta$(0)/{\it k$_{B}$T$_{c}$} = 3.9. Significantly, we observe a pseudogap in the superconducting state, corresponding to a suppression of intensity occurring over an energy scale up to the breathing mode phonon energy ( $\sim$ 70 meV ). This pseudogap is observed even above {\it T$_{c}$} but with a normal metallic Fermi edge. The pseudogap is filled up on increasing temperature in the normal phase. The results indicate electron phonon coupling induces a pseudogap in the metallic phase of Ba$_{0.67}$K$_{0.33}$BiO$_{3}$ . 

Single crystals of Ba$_{1-x}$K$_{x}$BiO$_{3}$, with x = 0.2, 0.33 and 0.46 were prepared by an electrochemical method and characterized as reported earlier[12]. Magnetization measurements on the superconducting sample used in the photoemission measurements confirmed T$_{c}$ = 31 K for x = 0.33 (see inset of Fig 1). Ultrahigh resolution angle-integrated photoemission spectra were measured using monochromatised He I$\alpha$ radiation from a GAMMADATA discharge lamp and a Scienta SES 2002 analyzer. The sample was cooled using a flowing liquid helium cryostat and the sample temperature was determined using a calibrated silicon diode sensor to an accuracy of  $\pm$0.5 K. Clean surfaces were obtained by scraping the sample surfaces using a diamond file. We also attempted cleaving but we always obtained uneven surfaces which showed no angle-dependence of spectra. We confirmed all the data on the scraped surfaces to be identical with the fractured surfaces. The total energy resolution is 4.2 meV as determined from the Fermi edge of a freshly evaporated gold film measured at 5.3 K. The position of the Fermi level(E$_{F}$) is accurate to better than $\pm$0.05 meV. Temperature dependent changes have been reproducibly obtained on thermal cycling.

Figure 1 shows the near E$_{F}$ spectra of Ba$_{0.67}$K$_{0.33}$BiO$_{3}$ obtained using He I$\alpha$ radiation at 5.3 K (superconducting phase) and 32 K (normal phase). The spectrum at 32 K (just above {\it T$_{c}$} = 31 K) is like that of a normal metal with a Fermi edge. This is confirmed by simulating the spectrum using a linear DOS multiplied by the Fermi-Dirac distribution and convoluted by a Gaussian of FWHM = 4.2 meV corresponding to the experimental resolution (see Fig. 1). In contrast, the spectrum at 5.3 K shows a sharp peak at 7.0 meV binding energy with a leading edge determined by the experimental resolution. The spectrum indicates a clear gap in the single particle DOS due to the superconducting transition. Using the Dynesf function[13] for the superconducting DOS, we have similarly simulated the spectrum at 5.3 K(Fig. 1) to obtain the gap value. We obtain a good fit for a gap value of $\Delta$(5.3 K) = 5.2 meV, and a $\Gamma$ value of 0.01 meV indicating negligible thermal smearing. While the fit deviates at binding energies beyond the peak, the good fit obtained for the leading edge and peak indicate that Ba$_{0.67}$K$_{0.33}$BiO$_{3}$ has an isotropic s-wave gap. The value of the gap gives a 2$\Delta$(0)/{\it k$_{B}$T$_{c}$} = 3.9, in agreement with tunneling[14] and optical spectroscopy experiments[15]. The estimated electron-phonon coupling parameter $\lambda$ $\sim$ 1.2$\pm$0.2 for x = 0.33 from tunneling[14] and $\lambda$ $\sim$ 1 from theory[16]. We have also calculated the photoemission spectrum at T = 5.3 K using Eliashberg equations in the strong coupling framework and we obtain a good maT$_{c}$h for a  $\lambda$ $\sim$ 1.2$\pm$0.2 and $\mu^{*}$ = 0.11, confirming the tunneling result. A 2$\Delta$(0)/{\it k$_{B}$T$_{c}$} = 3.9 implies that Ba$_{0.67}$K$_{0.33}$BiO$_{3}$ belongs to the moderately strong-coupling regime similar to niobium metal, which exhibits a 2$\Delta$(0)/{\it k$_{B}$T$_{c}$} = 3.8. 

The spectral weight redistribution shown in Fig. 1 is due to a systematic pile up in the DOS as shown in Fig. 2a. We have also simulated the superconducting spectra so as to obtain a measure of the temperature dependence of the gap.  As an example, we show the spectrum along with the simulation for T = 26 K, in the inset to Fig. 2b. Very interestingly, we see that the spectrum shows a small but clear peak above E$_{F}$ which is also well simulated and is due to the peak in the superconducting DOS above E$_{F}$. As the temperature increases, DOS just above the gap are populated, resulting in the peak above E$_{F}$. At the lowest temperature, we do not see this peak above E$_{F}$ due to the relatively large gap compared to the Fermi Dirac function(see Fig.1). The peak above E$_{F}$ grows in intensity and moves closer to E$_{F}$ with temperature and finally disappears above {\it T$_{c}$}. A similar peak above E$_{F}$ has been observed recently for V$_{3}$Si below T$_{c}$.[17] The gaps obtained from such fits to the spectra are plotted as a function of reduced temperature T/{\it T$_{c}$} in Fig. 2b. The plot matches well with the expected dependence for the gap as a function of T/{\it T$_{c}$} from BCS theory(shown as a line), which is similar to the strong coupling result[18]. 

In order to study the changes in the electronic structure associated with the semiconductor-metal transition we have measured the valence band spectra for Ba$_{1-x}$K$_{x}$BiO$_{3}$ (x = 0.2 and 0.33) with He I$\alpha$ radiation as shown in the inset to Fig. 3. The x = 0.2 spectrum was measured at 300 K as the sample charged up at low temperatures, while the x = 0.33 spectrum is measured at 5.3 K. The spectra are normalized at the most intense peak at 3.0 eV and, except for a small broadening of the peak at 3.0 eV, show little change in the gross features. Comparing with band structure calculations it is known that the most intense peak centred at 3.0 eV is due to non-bonding O 2p states, with the bonding Bi 6s-O 2p states at higher binding energy and the anti-bonding states occurring close to E$_{F}$[5]. The spectra are similar to that reported earlier using He II radiation but for the change in relative intensities due to a change in cross-sections consistent with the photon energy used for the measurements[19]. We concentrate on the Bi 6s-O 2p anti-bonding states, and as seen in Fig. 3, the spectrum for x = 0.2 does not show any intensity at E$_{F}$ due to the semiconducting nature of the sample. In comparison, the x = 0.33 spectrum shows remarkable changes associated with the increase in K substitution. As discussed above, the high-resolution low energy scale spectrum at 5.3 K shows a superconducting gap below {\it T$_{c}$} and a normal metallic Fermi edge above {\it T$_{c}$} . But most interestingly, at intermediate energy scales, we see a pseudogap at 5.3 K in the electronic structure occurring as a suppression in intensity upto 70 meV binding energy(Fig. 3). This energy corresponds to the highest phonon energy[20] and has also been associated with the breathing mode phonon energy ( 570 cm$^{-1}$ $\sim$ 70 meV ) as identified by Raman scattering[6]. 

We have investigated the temperature dependence of the pseudogap between 5.3 K and 300 K as shown in Fig. 4. While we observe a change in slope at about 70 meV in the 5.3 K and 32 K spectrum, the pseudogap is really not clear at high temperatures from a first look at the data. In order to get a clear picture of the changes as a function of temperature, we have normalized the spectra for the area under the curve and divided by a corresponding Fermi-Dirac function convoluted with the experimental resolution. We see (inset A to Fig. 4b) that the pseudogap surviving in the normal phase DOS over an energy of 70 meV gets filled up on increasing temperature to 300 K, without any changes in higher binding energies upto at least 300 meV. Since the spectral weight ought to be conserved, the absence of any spectral feature to compensate for the changes suggests spectral weight conservation over wider energy scales. The DOS are asymmetric with respect to E$_{F}$ as we limit the analysis to an energy of 3{it k$_{B}$}T above E$_{F}$ as determined by the Fermi Dirac function[21]. We have also measured the pseudogap and the superconducting transition in the overdoped BKBO with x = 0.46. The pseudogap is weaker in energy and is closed by T = 150 K as seen from the analysis shown in inset B to Fig.4. This suggests that the high energy phonons enhance the {\it T$_{c}$} and the 2$\Delta$(0)/{\it k$_{B}$T$_{c}$} to 3.9 in the optimal doping samples while for overdoping 2$\Delta$(0)/{\it k$_{B}$T$_{c}$} $\sim$ 3.5. We note that the pseudogap is a true pseudogap and not extrinsic as the pseudogap is more pronounced at low temperatures when the solid is more conducting, in contrast to the proposal by Joynt[22] for poorly conducting solids. In comparison, a normal phase pseudogap is also observed in the underdoped cuprates[2] but at low energy scales and develops into an anisotropic superconducting gap below {\it T$_{c}$}, although a higher energy scale pseudogap has also been discussed[23,24].

An effective attractive interaction between electrons (negative U Hubbard model) due to strong electron-phonon coupling[25,26], or due to an electronic origin[27,28] has been discussed in the context of the BPBO and BKBO series. Real-space on-site pairing in the insulating CDW state going over to a metallic phase with k-space pairing was proposed for the BPBO series[25]. The existence of a pseudogap existing over an energy scale of the CDW gap energy has also been calculated for the BPBO series[25,26]. eMissingf spectral weight identified in optical spectroscopy[15 and references therein] has been similarly discussed in the framework of a bipolaronic model. However, for the high-{\it T$_{c}$} cuprates, the existence of pairing and a spin-gap in the normal phase of the 2D negative-U Hubbard model has been shown to be consistent with experiments[29]. Further, it also exhibits a pseudogap in the single-particle DOS. The pseudogap we observe could also have a similar origin in pairing due to the fact that magnetic susceptibility also shows a gradual increase with temperature in the normal phase[12]. However, in a model for the high-{\it T$_{c}$} cuprates[30], it has been shown that preformed pairs can form at a temperature T$^{*}$ above {\it T$_{c}$} with phase coherence setting in at {\it T$_{c}$}. The existence of two temperature scales is in contrast to BCS behavior and may not be consistent with the BCS-like behavior of the superconducting gap seen here. Alternatively, it could be due to remnant local CDW order surviving in the metallic phase as observed in optical measurements and is consistent with the observation that the long and short Bi-O bonds become indistinguishable around x = 0.4 in EXAFS measurements. Given the proximity of the CDW state to superconductivity, it is possible that the BKBO series manifests electron-phonon coupling induced electron pairing above {\it T$_{c}$} as an intermediary metallic phase in Ba$_{0.67}$K$_{0.33}$BiO$_{3}$. The seemingly competing behavior of CDW order (or real-space pairing) and superconductivity (or k-space pairing) could then be derived from this intermediate phase by changes in electron-phonon coupling, carrier density and temperature. 

In conclusion, the superconducting gap in Ba$_{0.67}$K$_{0.33}$BiO$_{3}$ opens below {\it T$_{c}$} with a pile-up resulting in a sharp peak in the DOS at low energy scales. From 2$\Delta$(0)/{\it k$_{B}$T$_{c}$} = 3.9, Ba$_{0.67}$K$_{0.33}$BiO$_{3}$ belongs to the moderately strong electron-phonon coupling regime. A pseudogap is observed below and above {\it T$_{c}$}, occurring as a suppression in intensity up to the highest phonon energy($\sim$ 70 meV), also associated with the breathing mode phonon. The pseudogap fills up on increasing temperature in the normal phase. The results indicate electron-phonon coupling induces a pseudogap in Ba$_{0.67}$K$_{0.33}$BiO$_{3}$ .

Acknowledgements : We thank Professors M. Imada and M. Takigawa for very valuable discussions. This work was supported by grants from the Ministry of Education, Science, Sports and Culture of Japan. 


\newpage

\begin{figure}
\centerline{\epsfig{file=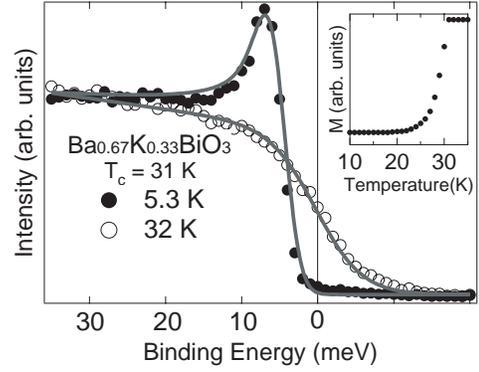,width=7cm}}
\hspace{0.5cm}
\caption{Fig. 1. Ultrahigh resolution photoemission spectra of Ba$_{0.67}$K$_{0.33}$BiO$_{3}$ near E$_{F}$ (photon source : He I$\alpha$, h$\nu$ = 21.218 eV) at 5.3 and 32 K. The 5.3 K spectrum shows a clear superconducting gap. The lines are fits to the normal and superconducting states (see text). Inset : Magnetization vs. T.}
\label{fig1}
\end{figure}

\begin{figure}
\centerline{\epsfig{file=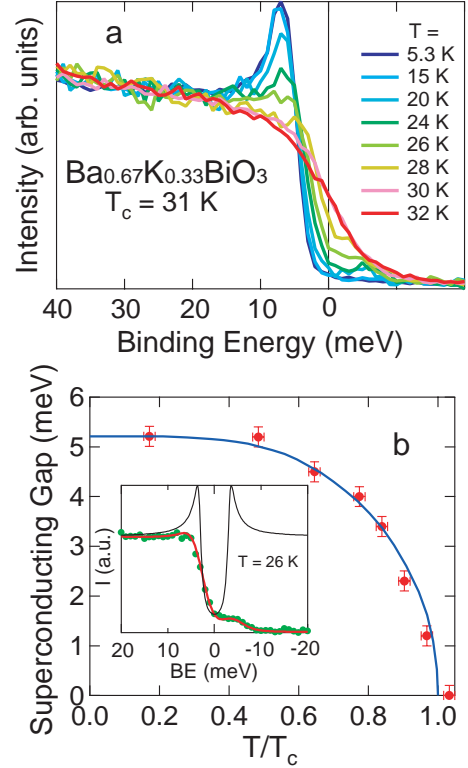,width=6.8cm}}
\hspace{0.5cm}
\caption{Fig. 2a. The temperature dependent spectra exhibit a systematic pile up in the DOS. Fig. 2b. Experimental (symbols) and BCS (line) gap vs. T/{\it T$_{c}$}. Inset to Fig. 2b : the spectrum (symbols) with the fit (red line) for T = 26 K. The small peak above E$_{F}$ originates in the superconducting DOS (black line). }
\label{fig2}
\end{figure}

\begin{figure}
\centerline{\epsfig{file=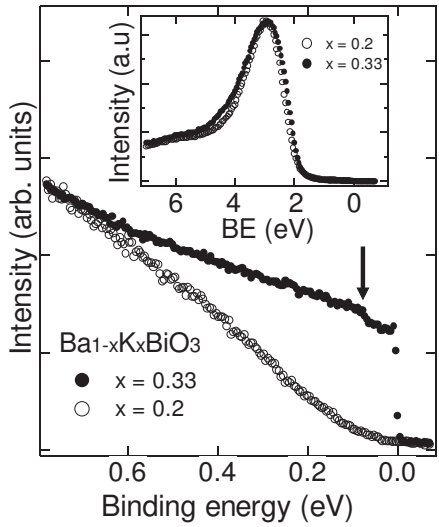,width=7cm}}
\hspace{0.5cm}
\caption{Fig. 3. The valence band spectra of Ba$_{1-x}$K$_{x}$BiO$_{3}$ (x = 0.2 and 0.33) showing  changes across the semiconductor-metal transition and a pseudogap over  $\sim$ 70 meV(arrow) for x = 0.33 at 5.3 K. The superconducting transition is not clear here due to the larger step size used. Inset shows the full valence band spectra.
}
\label{fig3}
\end{figure}

\begin{figure}
\centerline{\epsfig{file=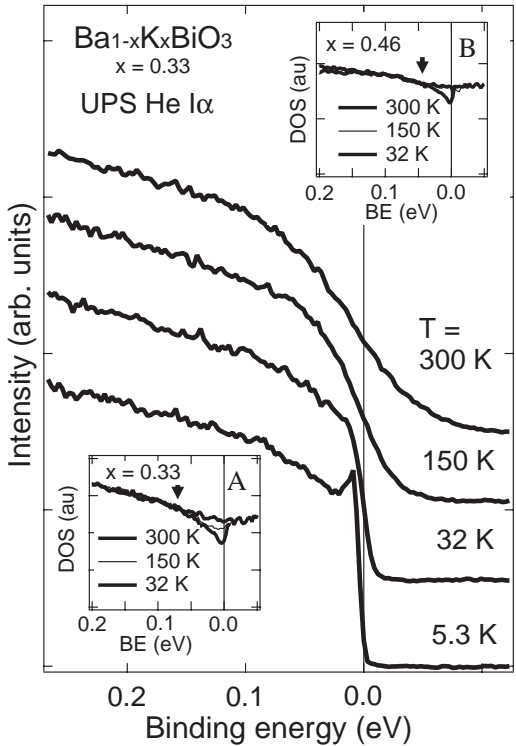,width=7cm}}
\hspace{0.5cm}
\caption{Fig. 4. Temperature dependent Ba$_{0.67}$K$_{0.33}$BiO$_{3}$ spectra between 5.3 and 300 K. Insets shows the pseudogap energy is fixed (arrow) for a particular x ; it gets filled up by 300 K for x = 0.33(inset A) and 150K for x = 0.46(inset B)
}
\label{fig4}
\end{figure}
\end{document}